\documentclass[%
reprint,
amsmath,amssymb,
aps,
prd,
floatfix,
]{revtex4-2}

\usepackage{graphicx}
\usepackage{dcolumn}
\usepackage{bm}
\usepackage[colorlinks=true,citecolor=blue,linkcolor=blue,urlcolor=blue]{hyperref}

\usepackage{tikz}
\usepackage{xcolor}
\usepackage{soul}

\newcommand{\orcidicon}[1]{\href{https://orcid.org/#1}{\includegraphics[height=\fontcharht\font`\B]{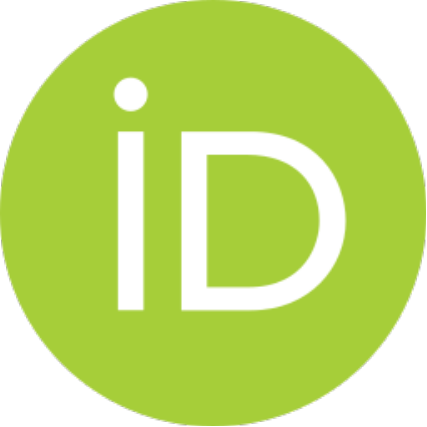}}}

\usepackage{mathrsfs}
\usepackage{pstricks}
\usepackage{color}
\usepackage{slashed}
\usepackage{epsfig}
\usepackage{amsfonts}
\def\beq{\begin{equation}}
\def\eeq{\end{equation}}
\def\bea{\begin{eqnarray}}
\def\eea{\end{eqnarray}}

\def\Re{\textrm{Re}}

\def\Tr{\textrm{Tr}}

\def\d{\mathrm{d}}

\def\q_perp{\mathbf{q}_{\perp}}
\RequirePackage[normalem]{ulem}
\RequirePackage{color}

\newcommand{\inner}[2]{\langle #1,#2\rangle} 

\newcommand{\norm}[1]{\|{#1}\|}

\newcommand{\R}{\mathbb{R}}

\begin{document}

\title{
Bounds in partition functions of the continuous \\ random field Ising model}
\author{G. O. Heymans\,\orcidicon{0000-0002-1650-4903 }}
\email{Email address: olegario@cbpf.br}
\affiliation{Centro Brasileiro de Pesquisas F\'{\i}sicas - CBPF, \\ Rua Dr. Xavier Sigaud 150 22290-180 Rio de Janeiro, RJ, Brazil}
\author{N.~F.~Svaiter\,\orcidicon{0000-0001-8830-6925}}
\email{Email address: nfuxsvai@cbpf.br}
\affiliation{Centro Brasileiro de Pesquisas F\'{\i}sicas - CBPF, \\ Rua Dr. Xavier Sigaud 150 22290-180 Rio de Janeiro, RJ, Brazil}

\author{B.~F.~Svaiter\,\orcidicon{0000-0003-3656-9883}}
\email{Email address: benar@impa.br}
\affiliation{Instituto de Matemática Pura e Aplicada - IMPA \\ Estrada Dona Castorina 110 22460-320 Rio de Janeiro, RJ, Brazil}

\author{A. M. S. Macêdo\,\orcidicon{0000-0002-4522-031X}}
\email{Email address: antonio.smacedo@ufpe.br}
\affiliation{Universidade Federal de Pernambuco - UFPE \\ Av. Prof. Moraes Rego 1235 50670-901, Recife, PE, Brazil}


\begin{abstract}
We investigate the critical properties of continuous random field Ising model (RFIM).  Using the distributional zeta-function method, we obtain a series representation for the quenched free energy. It is possible to show that for each moment of the partition function, the multiplet of $k$-fields the Gaussian contribution has one field with the contribution of the disorder and $(k-1)$-fields with the usual propagator. Although the non-gaussian contribution is non-perturbative we are able to show that the model is confined between two $\mathbb{Z}_2\times\mathcal{O}(k-1)$-symmetric models. Using arguments of lower critical dimension alongside with monotone operators, we show that the phase of the continuous RFIM can be restricted by an $\mathbb{Z}_2 \times \mathcal{O}(k-1) \to \mathcal{O}(k-2)$ phase transition.

\end{abstract}



\maketitle




\textit{Introduction} -- In the physics of disordered systems, 
the random field Ising model (RFIM) is under intensive theoretical \cite{nattermann1989random,young1998spin}, experimental and numerical studies (for recent works, see $e.g.$ \cite{fytas2016phase, Fytas:2016itl}). This model was introduced by Larkin, to study vortices in superconductors \cite{larkin1970effect}. Applying a uniform external field in a diluted Ising antiferromagnet, the RFIM can be realized in the laboratory \cite{fishman1979random}. Other well known example of experimental realization of the RFIM is binary liquids in random porous media \cite{de1984liquid}. The RFIM in a hypercubic  $d$-dimensional lattice is defined by the Hamiltonian 
\begin{equation}
H=-J\sum_{(i,j)=1}^{N}\,S_{i}S_{j}-\sum_{i=1}^N\,h_{i}S_{i}, \label{1}
\end{equation}
where $(i,j)$ indicates that the sum is taken over nearest neighbor pairs and $S_{i}=\pm 1$. 
In the above equation $N$ is the total number of Ising spins. Periodic boundary conditions can be used and the thermodynamic limit must be used in the end.
The partition function is $Z=\Tr\, e^{-\beta H}$. 
In  Eq. (\ref{1}) the $h_{i}$'s are the quenched random variables totally uncorrelated on different sites. The \emph{quenched} (Gibbs) free energy is defined as 
$F=\mathbb{E}[\ln Z]$ where $\mathbb{E}\,[...]$ means average over 
the ensemble of all realizations of the disorder.
Here  we consider  
a Gaussian distribution. 
%
%
The probability distribution of such quenched random variables has zero mean-value, $\mathbb{E}\,[h_{i}]=0$, and correlation functions given by
$\mathbb{E}\,[h_{i}h_{j}]=h_{0}^{2}\delta_{ij}$.

A central question in the physics of disordered systems is the comparison between critical behavior of systems in the absence and in the presence of disorder \cite{Harris:1974zug}. That is, what is it the nature of the transition of models from the symmetric to the ordered phase?
There are two dimensions that are of particular relevance 
in pure and  quenched disordered models.
The first one is the lower critical dimension $d_{c}^{\,-}$, which is the lowest value of the spatial dimension where there is no long range order.
The second one is the upper critical dimension $d_{c}^{\,+}$, the dimension above which the model is Gaussian in the infrared.
In Ref. \cite{Imry:1975zz} the following  two results were presented for the RFIM, which is dominated by disorder fluctuations.
Using Pierls arguments \cite{peierls1936ising}, these authors proved that $d_{c}^{\,-}=2$.
Using renormalization group techniques, they also proved that $d_{c}^{\,+}=6$.
The first result was discussed by Imbrie \cite{Imbrie:1984ki} and the latter confirmed by Aizenman and Wehr \cite{Aizenman:1989dk, Aizenman:1990ji}.
Concerning the existence of the phase transition, in Refs. \cite{Bricmont:1987fv, bricmont1988phase}, it was proved that there is an ordered phase for $d\geq 3$. See also the Ref. \cite{grinstein1984lower}.
Other important result discussing the behavior of the pure and the disordered model were obtained by many authors. It was proved that the critical exponents of the system with quenched disorder are identical to the critical exponents of the pure system in $(d-2)$ dimensions \cite{grinstein1976ferromagnetic, Aharony:1976jx, Parisi:1979ka, Kaviraj:2021qii, Cardy:2023zna,klein1984supersymmetry}.
%
%

The intriguing interplay between disorder and criticality in the RFIM, especially the dimensional reduction property, provides a fertile ground for exploring its applications beyond its original formulation. Recent studies on quasi-long-range order in the RFIM reveal how local Markov statistics can give rise to intermediate states between long-range order and disorder. This emergence of complex behavior from simple local rules resonates with phenomena observed in diverse fields where Markov random fields find application, such as image analysis, network science, and even social dynamics \cite{Enrique}. 

The soft version of the RFIM is a Landau-Ginzburg model with an additive quenched disorder. The aim of this paper is to study the critical properties of Landau-Ginzburg model in the presence of additive quenched disorder, i.e., a non-thermal control parameter.
%
%
%
In this setting, one 
can find the average free energy  \cite{englert1963linked, griffiths1968random}.
%
One way to find such an average is the replica trick \cite{edwards1975theory, Emery:1975zz,mezard1987spin, dotsenko2005introduction}. %
Other  possibilities discussed the dynamics of systems with quenched disorder \cite{de1978dynamics,sompolinsky1981dynamic,de2006random} and the other uses Grassmann anticomuting variables \cite{Efetov:1983xg}.
An alternative method discussed in the literature is the distributional zeta-function method \cite{Svaiter:2016lha,Svaiter:2016-2, Diaz:2017grg,Diaz:2017ilf, Diaz:2016mto}, which is adopted in this work. 


%
%

Using the distributional zeta-function method and tools of functional analysis, here we obtain a new result. Instead of show that the critical behavior of the model is described by the critical behavior of a system without disorder in a reduced effective dimension, here we show that the phase transition of the model must be bounded by phases transitions of the kind $\mathbb{Z}_2 \times \mathcal{O}(k-1) \to \mathcal{O}(k-2)$.

Let us briefly review the functional formalism for pure systems, that is, systems without disorder \cite{Wilson:1973jj}.
In such a scenario (pure) Ising Hamiltonian is replaced by
Landau-Ginzburg functional, which is defined as
\begin{equation}
S(\varphi)=
\int d^{d}x\left[\frac{1}{2}\varphi(x)\left(-\Delta+m_{0}^{2}\right)\varphi(x)+\frac{\lambda_{0}}{4!}\varphi^{4}(x) \right],
\label{9}
\end{equation}
%
where $\Delta$ denotes the Laplacian in 
$\mathbb{R}^{d}$, $\lambda_{0}$ is the bare coupling constant, and $m_{0}$ is a spectral parameter of the model, which will be called by mass.  
The functional integral
\begin{equation}\label{eq:part}
Z=\int_{\partial\Omega} [d\varphi]\,\,\exp\bigl(-S(\varphi)\bigr),
\end{equation}
defines the partition function, where $[d\varphi]$ is a functional measure, given by $[d\varphi]=\prod_{x} d\varphi(x)$, and $\partial\Omega$ in the functional integral means that the field $\varphi(x)$ satisfies some boundary condition in the boundary $\partial\Omega$ of the domain. The thermodynamic limit (infinite volume) must be assumed in the end.
To preserve translational invariance, periodic boundary conditions can be imposed replacing $\mathbb{R}^{d}$ by the torus $\mathbb{T}^{d}$. 
With the partition function, one can construct a probability measure equivalent to the Gibbs measure in statistical mechanics. Therefore, the average value for any polynomial of the field $f(\varphi)$ is given by
\begin{equation}
\langle f(\varphi) \rangle =\frac{1}{Z}\int [d\varphi]f(\varphi) \exp\bigl(-S(\varphi)\bigr).
\end{equation} 
With the above definition, all the $n$-point correlation functions of the model can be found.
Introducing an external source $j(x)$, one can define the generating functional of all $n$-point correlation functions $Z(j)$ as 
%
\begin{equation}
Z(j)=\int_{\partial\Omega} [d\varphi]\,\, \exp\left(-S(\varphi)+\int d^{d}x j(x)\varphi(x)\right).
\label{eq:generatingfunctional}
\end{equation}
Next, using the linked cluster theorem, it is possible to define the generating functional of connected correlation functions, given by $W(j)=\ln Z(j)$. Both generating functionals can be represented by  Volterra series. 
Taking functional derivatives with respect to the external source and setting to zero in the end, we obtain the $n$-point correlations functions and the $n$-point connected correlations functions of the model, respectively.

To discuss disordered systems, let us introduce the functional $Z(j,h)$, the generating functional of correlation functions for one disorder realization, where we again use a external source $j(x)$. This functional integral is defined
\begin{equation}
Z(j,h)=\int_{\partial\Omega} [d\varphi]\,\, \exp\left(-S(\varphi,h)+\int \d^{d}x\, j(x)\varphi(x)\right),
\label{eq:disorderedgeneratingfunctional}
\end{equation}
where the action functional in the presence of disorder is defined as 
\begin{equation} 
S(\varphi,h)=S(\varphi)+ \int \d^{d}x\,h(x)\varphi(x).
\end{equation}
In the above equation, $S(\varphi)$ is the pure Landau-Ginzburg action functional, defined in Eq. (\ref{9}) and $h(x)$ is 
a quenched disorder

To perform such a disorder averages,
one defines for one disorder realization $\ln Z(h,j)$ and after it, one computes the average over all disorder realizations. 
As in the pure system case, one can define 
the generating functional of connected correlation functions for one disorder realization, $W(j,h)=\ln Z(j,h)$. 
Therefore, we define the disorder average of the $W(j,h)$, i.e., the quenched free energy. This generating functional is written as
\begin{equation}
\mathbb{E}\bigl[W(j,h)\bigr]=\int\,[\d h]P(h)\ln Z(j,h).
\label{eq:disorderedfreeenergy}
\end{equation} 
The conventional way to obtain the correlation functions is taking the functional derivative of $ \mathbb{E}\bigl[W(j,h)\bigr])$ with respect to $j(x)$ directly using the Eq. \eqref{eq:disorderedfreeenergy}.
Note that to obtain all the correlation functions of the model, one has to deal with the contribution $(Z(j,h))^{-1}$. Notice that although we have averaged over the disorder obtaining a model without spatial heterogeneities, the effects of local fluctuation of the disorder in the original system will appear in this formalism.

\textit{Distributional zeta-function} -- For a general disorder probability distribution, using the disordered functional integral $Z(j,h)$ given by Eq. \eqref{eq:disorderedgeneratingfunctional}, the distributional zeta-function, $\Psi(s)$, is defined as
\begin{equation}
\Psi(s)=\int [dh]P(h)\frac{1}{Z(j,h)^{s}},
\label{pro1}
\vspace{.2cm}
\end{equation}
\noindent for $s\in \mathbb{C}$, this function being defined in the region where the above integral converges. The average generating functional can be written as 
\begin{equation}
\mathbb{E}\bigl[W(j,h)\bigr]=(d/ds)\Psi(s)|_{s=0^{+}}, \,\,\,\,\,\,\,\,\,\, \Re(s) \geq 0,  
\end{equation}
where one defines the complex exponential $n^{-s}=\exp(-s\log n)$, with $\log n\in\mathbb{R}$.
Using analytic tools, the average free energy can be represented as
\begin{align}\label{eq:zeta}
\mathbb{E}\left[W(j,h)\right]&=\sum_{k=1}^{\infty} \frac{(-1)^{k+1}c^k}{k k!}\,\mathbb{E}\left[Z^k(j,h)\right]\nonumber  \\
&-\ln(c)+\gamma+R(c,j)
\end{align}
where the quantity $c$ is a dimensionless arbitrary constant, $\gamma$ is the Euler-Mascheroni constant, and $R(c)$ is given by
\begin{equation}
R(c,j)=-\int [dh]P(h)\int_{c}^{\infty}\,\dfrac{dt}{t}\, e^{-Z(j,h)t}.
\end{equation}
\noindent Therefore for large $c$, $|R(c)|$ is quite small, therefore, the dominant contribution to the quenched free energy is given by the moments of partition functions of the model.  
%
%
We are using a Gaussian disorder, i.e., $\mathbb{E}[{h(x)h(y)}]=\sigma^{2}\delta^{d}(x-y)$. To discuss the ordered phase in the model, i.e., the infrared regime, one expects that  the microscopic details of the disorder must be irrelevant.  After integrating over the disorder we get that each moment of the partition function $\mathbb{E}\,[Z^k(j,h)]$ can be written as
\begin{equation}
\mathbb{E}\left[Z^k(j,h)\right]=\int\,\prod_{i=1}^k\left[\d\varphi_{i}^{(k)}\right]\,\exp\left(-S_{\textrm{eff}}^{(k)}\left(\varphi_{i}^{(k)},j_{i}^{(k)}\right)\right),
\label{aa11}
\end{equation}
\noindent where the effective action $S_{\textrm{eff}}^{(k)}\left(\varphi_{i}^{(k)}\right)$ describing the field theory with $k$-field components. For now on, we are omitting the super-index $k$ in the fields variables; in this new notation the effective action reads
\begin{widetext}
\begin{equation}\label{eq:Seff1}
S_{\textrm{eff}}^{(k)}(\varphi_{i},j_{i})= \int \d^{\,d}x\left[\sum_{i=1}^{k}\left(\frac{1}{2}\,\varphi_{i}(x)\left(-\Delta+m_{0}^{2}\right)\varphi_{i}(x) +\frac{\lambda_{0}}{4!}(\varphi_{i}(x))^{4}\right)-\frac{\sigma^{2}}{2}\sum_{i,j=1}^{k}\varphi_{i}(x)\varphi_{j}(x)-\sum_{i=1}^{k}\varphi_{i}(x)j_{i}(x)\right].
\end{equation}
\end{widetext}
%

\textit{Free theory} -- In view of Eq.(\ref{eq:Seff1}), the propagator of our effective theory is a $k\times k$ full-matrix, i.e., is a non-diagonal propagator. Such feature have been explored in the literature in different ways ~\cite{Lewandowski:2017omt,Lewandowski:2018bnn, Heymans:2023tgi, Heymans:2024dzq}. In the context of the distributional zeta function, two ways have been presented. First, one make a diagonal ansatz in the function space: $\varphi_i = \varphi_j,\,\, \forall \,\,i,j$. With such ansatz the perturbation theory can be carried out in the usual way and consistent results have been found for both, random field and random mass cases \cite{Diaz:2016mto, Diaz:2017grg, Diaz:2017ilf, Soares:2019fed, Rodriguez-Camargo:2021ryf, Rodriguez-Camargo:2022wyz, Heymans:2022sdr}. For instance, using the diagonal ansatz one can show that we recovered the correct upper critical dimension for the RFIM, $d_c^+ = 6$ \cite{binney1992theory}. The question that arises is what are the results we can find without assuming the diagonal ansatz. With that in mind, was proposed the second way, the diagonalization method. This approach emerge once one analyses the free-part of the effective action
\begin{equation}
    \sum_{i,j=1}^kS_0(\varphi_i, \varphi_j) = \frac{1}{2} \sum_{i,j=1}^k\int \d^d x
    \varphi_i(x) \left(G^0_{ij} - \sigma^2\right)\varphi_j(x),
\end{equation}
where $G^0_{ij} =  \left(-\Delta + m_0^2\right)\delta_{ij}$. 
Such an action can be equivalently represented by
\begin{equation}\label{eq:free}
        \sum_{i,j=1}^k S_0 (\varphi_i, \varphi_j) = \frac{1}{2}\int \d^dx \,\, \inner{\Phi}{G\Phi},
\end{equation}
where $G$ is the $k\times k$ full matrix with components $G^0_{ij} - \sigma^2$, $\Phi(x)$ is the vector with components $\varphi_i(x)$, and $\inner{\cdot}{\cdot}$ is the natural inner product in $\R^k$. Now noticing that $G$ is real and symmetric, one can find its diagonalization by an orthogonal matrix $O$:
\begin{equation}\label{eq:diag}
    D = \langle O, GO\rangle = \left[
\begin{array}{cccc}
  G^0_{11} -k\sigma^2 & 0& \cdots & 0 \\
   0 & G^0_{22} & \cdots & 0 \\
   \vdots & \cdots & \ddots & \vdots \\
   0 & \cdots & & G^0_{kk}
\end{array}
\right]_{k\times k}.
\end{equation}

Foremost we should notice that, from the start, in $\mathbb{R}^k$, which appears as a result of the average, does not have any special properties. Besides the usual vector space properties, the Eq. (\ref{eq:Seff1}) does not impose any other qualities in this space. Then, to keep the formulation general as possible, we shall assume minimal properties over $\mathbb{R}^k$.
Now, defining that $\tilde{\Phi}(x) = O\Phi(x)$ is the vector with components $\tilde{\Phi} = (\phi,\phi_1,\dots, \phi_{k-1})$, we are able to present a third expression of the free effective action
\begin{align}
    &\sum_{i,j=1}^k S_0 (\phi_i) = \frac{1}{2}\int \d^dx \,\, 
      \phi(x)(-\Delta + m_0^2 - k\sigma^2) \phi(x) \nonumber \\
      &\hspace{1.65cm} +   \frac{1}{2} \sum_{a=1}^{k-1} \int \d^dx \,\, 
      \phi_a(x)(-\Delta + m_0^2) \phi_a(x),
\end{align}
which is clearly the sum of $k$ free actions with two distinct differential operators. 

As we have seemed, there is no problem in the application of the diagonalization approach, Eq. (\ref{eq:diag}), for the free effective action. The functional measure is also well-behaved under the diagonalization, once that the matrix which do the transformation is orthogonal the absolute value of Jacobian will be the unity. The source $j_i$, introduced to generate the correlation functions, can always be chosen in which way they transform with the inverse transformation of the vector $\Phi$, it is also well-behaved. From now on, we discuss the source-free case. So, for free actions, the diagonalization approach is able to describe the system without any ansatz over functional space.  A problem emerges once we turn on the interaction. 

\textit{Interacting theory} -- 
From Eq. (\ref{eq:diag}), there is always a set of $k-1$ degenerated eigenvalues. Which means that one needs to orthogonalize the respective eigenvectors, which are columns of $O$. 
This feature of the matrix $O$ inserts difficulties in the interacting part. As one can see from Eq. (\ref{eq:Seff1}), after the disorder average, the effective interaction is not symmetric by rotations in $\R^k$. Such an interaction is known in the literature as cubic anisotropic interaction \cite{aharony1973critical, Nattermann:1977uz, kleinert2001critical}. 

Technical difficulties arise when $k$ increases. Such a feature can be directly related to the non-pertubative behavior of the RFIM. However, here the non-pertubative behavior is of a different kind of the usual that appears in field theories. It is non-pertubative due to the impossibility to write explicitly the interaction for any value of $k$ after taking the quenched average. This situation is similar to the case of Bose-Hubbard model \cite{sachdev1999quantum}.

Nevertheless, we show that the effective action given by Eq.  (\ref{eq:Seff1}) has an upper and a lower bound which are rotational-symmetric. We will construct two effective actions in which the diagonalization procedure does not affect the interacting part, and such actions will establish an upper and a lower bound for the partition function of RFIM.

\textit{Bounds in partition functions--} Once that the free case have been treated and presents no problems, let us focus on the cubic anisotropic interaction
\begin{equation}
    S_{\mathrm{CA}}^{(k)}(\varphi_i) = \frac{\lambda_{0}}{4!}\int \d^dx\sum_{i=1}^{k}(\varphi_{i}(x))^{4}.
\end{equation}
We adopt the notation $ \norm{ \,\cdot\, }_p $ for the $ p $-norm in
$ \R^k$, so that
$\Vert \Phi ( x )\Vert_p = \left[\sum_i |\varphi_i ( x )|^p \right]^{1/p}$
for any $ x \in \R^d$;
hence, the interaction can be recast as
\begin{equation}
     S_{\mathrm{CA}}^{(k)}(\Phi) = \frac{\lambda_0}{4!}\int \d^dx\Vert \Phi (x) \Vert_4^4.
\end{equation}
With that in mind, we can go further.
Observe that for any $ a \in \R^k $,
\begin{equation}
  \norm{a}_1 \leq \sqrt{k} \;\; \norm{a}_2, \quad
  \norm{a}_2 \leq \norm{a}_1.
\end{equation}
The first above inequality can be proved writing
$ \norm{a} = \inner{a}{s}$ with $s_i=1$ if $a_i \geq 0$, $s_i=-1$ otherwise, and applying Cauchy-Schwarz inequality.
The second inequality can be verified by direct computation of
$ \norm{a}_1^2 - \norm{a}_2^2 $.
Fix $ x \in \R^d $ and set $a_i=\varphi_i(x)^2$  with
 for $i=1,\dots,k$ and
$ a = (a_1,\dots,a_k)$.
Since $\norm{a}_1 =   \norm{\Phi(x)}_2^2 $ and $ \norm{a}_2 =   \norm{\Phi(x)}_4^2$,
it follows from the above inequalities that
\begin{equation}
  \dfrac{\norm{\Phi(x)}_2^2}{\sqrt{k}}
   \leq
  \norm{\Phi(x)}_4^2
   \leq
  \norm{\Phi(x)}_2^2.
\end{equation}
%
%
This inequality can be used to obtain a bound for the cubic anisotropic interaction:
\begin{equation}
    \frac{1}{k}S_{\mathcal{O}(k)}(\Phi) \leq S_{\textrm{CA}}^{(k)}(\Phi) \leq S_{\mathcal{O}(k)}(\Phi),
\end{equation}
where we have defined the interaction action
\begin{equation}\label{eq:ok}
     S_{\mathcal{O}(k)}(\Phi) = \frac{\lambda_{0}}{4!}\int \d^dx\norm{\Phi(x)}_2^4.
\end{equation}

Such a result is useful once that, for all $x  \in  \R^d$, the norm $\norm{\, \cdot \,}_2$ is invariant under orthogonal transformations in $\R^k$:
\begin{align}
    &\norm{ \Phi (x)}_2 = \sqrt{\inner{\Phi(x)}{\Phi(x)}} = \sqrt{\inner{ O^T\tilde{\Phi}(x)}{O^T\tilde{\Phi}(x)}} \nonumber \\
    &= \sqrt{\inner{\tilde{\Phi}(x)}{O O^T\tilde{\Phi}(x)}}= \sqrt{\inner{\tilde{\Phi}(x)}{\tilde{\Phi}(x)}} =  \norm{ \tilde{\Phi}(x) }_2,
\end{align}
for any $O$ orthogonal matrix, that is $O^T O = I$, with $O^T$ denoting the transpose of $O$. From that, one can use the actions given by Eqs. (\ref{eq:free}) and (\ref{eq:ok}) to construct the following actions
\begin{align}
    &S_{\textrm{U}}^{(k)}(\Phi) = \frac{1}{2}\int \d^dx \,\, \inner{\Phi(x)}{G \Phi(x)} + \frac{\lambda_{0}}{4!}\int \d^dx\norm{ \Phi (x)}_2^4, \\
    &S_{\textrm{L}}^{(k)}(\Phi) = \frac{1}{2}\int \d^dx \,\, \inner{\Phi(x)}{G \Phi(x)} + \frac{\lambda_{0}}{4!k}\int \d^dx\norm{ \Phi (x)}_2^4,
\end{align}
These actions are natural upper and lower limits to the effective action given by Eq. (\ref{eq:Seff1}), that is,
\begin{equation}
    S_{\textrm{L}}^{(k)}(\Phi) \leq S_{\textrm{eff}}^{(k)}(\Phi) \leq S_{\textrm{U}}^{(k)}(\Phi),
\end{equation}
and also, due to the property of the norm $\norm{\,\cdot\,}_2$, both have nicer orthogonal transformations on $\R^k$.

Using the same orthogonal matrix that have been used to diagonalize $G$, Eq. (\ref{eq:diag}), we can write the diagonalized action in terms of the components of $\tilde{\Phi}$ as
\begin{widetext}
\begin{equation}\label{eq:acbound}
    S_{\#}^{(k)}(\phi,\phi_a) =\int \d^dx\left[ \frac{1}{2} \phi(x)(-\Delta + m_0^2 - k\sigma^2) \phi(x) +  \frac{1}{2} \sum_{a=1}^{k-1} \phi_a(x)(-\Delta + m_0^2) \phi_a(x) + \frac{\lambda_{\#}}{4!} \left( \phi^2(x) + \sum_{a=1}^{k-1}\phi_a^2(x)\right)^2\right],
\end{equation}
\end{widetext}
with $S_{\#}^{(k)} =  S_{\textrm{U}}^{(k)},$ and $S_{\textrm{L}}^{(k)}$, adopting $\lambda_{\#}= \lambda_0,$ and $\lambda_0/k$ respectively. Analyzing such an action, we can verify that it is the action for two different kinds of scalar fields, with different masses. The underlying symmetry of this action is $\mathbb{Z}_2 \times \mathcal{O}(k-1)$. In different contexts, such actions have been studied \cite{Chai:2020zgq, Heymans:2022idr}. One interesting feature is that, considering any phase transitions, such an action intrinsically preserves the no-go theorems of  Mermin-Wagner, Hohenberg, and Coleman~\cite{Mermin:1966fe,Hohenberg:1967zz, Coleman:1973ci}. 

Now we can construct the partition function for each of these actions. Due to the monotonicity of the exponential, we get that
\begin{equation}
    Z_{\textrm{L}}^{(k)} \leq \mathbb{E}\left[Z^k(j,h)\right] \leq Z_{\textrm{U}}^{(k)},
\end{equation}
where
\begin{align}
    &Z_{\textrm{L}}^{(k)} = \int\prod_{i=1}^k[\d \varphi_{i}]\exp\left(-S_{\textrm{U}}^{(k)}(\Phi)\right),\label{eq:low} \\
    &Z_{\textrm{U}}^{(k)} = \int\prod_{i=1}^k[\d \varphi_{i}]\exp\left(-S_{\textrm{L}}^{(k)}(\Phi)\right).\label{eq:up}
\end{align}
That is, without any ad hoc choice of subsets in $\R^k$, we are able to obtain partition functions that are bounds, in each term of the series of Eq. (\ref{eq:zeta}). 

\textit{Conclusions} -- This is our main result: the effective action obtained by taking the average over the quenched disorder, which presents a non-perturbative behavior, can be confined between two pertubatively well-behaved actions. 
The fundamental question that can be answered with the results of this letter is the nature of the phase transition of the continuous RFIM.
This problem can be examined using the concepts of the lower critical dimension of the RFIM and no-go theorems. The model is bounded by two theories: $\mathbb{Z}_2 \times \mathcal{O}(k-1)$.
In these theories there are two different phase transitions: (i) $\mathbb{Z}_2 \times \mathcal{O}(k-1) \to \mathcal{O}(k-1)$, and (ii) $\mathbb{Z}_2 \times \mathcal{O}(k-1) \to \mathcal{O}(k-2)$ \cite{Heymans:2022idr}. Since the lower critical dimension for the RFIM is two, the case (i) cannot represent a phase transition due to disorder. 
We have thus obtained a new result. The phase transition of the continuous RFIM can be restricted by a $\mathbb{Z}_2 \times \mathcal{O}(k-1) \to \mathcal{O}(k-2)$ phase transition. The situation is similar to that of the cubic anisotropic model, which is confined between the Ising model and the Heisenberg model. A question that can be answered by this method is that if the nature of the phase transition depends on the particular choice of the probability distribution for the random field \cite{fytas2013critical, fytas2018review}. In the case of a different probability distribution, the symmetry of the bounds can shed some light in such a question. Also, it is possible to elucidate the nature of the phase transition in the continuous RFIM by connecting the bounds established in the distributional zeta-function approach using the interpolation method similar to the one pioneered by Guerra \cite{Guerra2003}. A interpolation between the diagonal ansatz and the bounds can be made as follows. Noting that the interaction in the diagonal approximation resembles the self-interaction of the field variable $\phi(x)$ in Eq. (\ref{eq:acbound}) we may define a new field variable $\tilde{\psi}(x) = A \tilde{\Phi}(x)$, with $A_{11} = 1$, ${A_{ab} = \sqrt{t}\delta_{ab}}$, $A_{1a} = A_{a1}=0$  for $a,b \in \{2,\dots, k\}$, and $t \in [0,1]$, so that the new action interpolates between the bounds and the diagonal ansatz. The explicit calculations of the critical exponents of the upper and lower $\mathbb{Z}_2 \times \mathcal{O}(k-1)$ theories, which can be straightforwardly done using the methods of \cite{parisi1998statistical}, along with a similar calculation for the $t$-dependent action, would then lead to strong statements about the nature of the phase transition in the continuous RFIM, including the intriguing problem of dimensional reduction. We conclude by remarking that the effective actions derived here provide a robust framework for this analysis.


\section*{Acknowledgments} 
The authors are grateful to S. A. Dias and G. Krein for fruitful discussions. This work was partially supported by Conselho Nacional de Desenvolvimento Cient\'{\i}fico e Tecnol\'{o}gico (CNPq), grants nos. 303436/2015-8 (N.F.S.), 311300/2020-0 (B.F.S.), and 307626/2022-9 (A.M.S.M.), and Fundação Carlos Chagas Filho de Amparo à Pesquisa do Estado do Rio de Janeiro (FAPERJ) grant no. E-26/203.318/2017 (B.F.S.). G.O.H. thanks to Fundação Carlos Chagas Filho de Amparo à Pesquisa do Estado do Rio de Janeiro (FAPERJ) due the Ph.D. scholarship.  


%

\end{document}